\documentclass[amsmath,amssymb,twocolumn,prd,floatfix,showpacs, nofootinbib]{revtex4-2}
\usepackage{tabularx}
\usepackage{bm, color}
\usepackage{overpic,subfigure}
\usepackage{multirow}
\usepackage{array}
\usepackage{dcolumn}
\usepackage[symbol]{footmisc}
\usepackage{epstopdf}
\usepackage[normalem]{ulem}
\usepackage{booktabs}
\usepackage{hyphenat}
\usepackage{makecell}

\RequirePackage{xspace}
\newcommand{\gev}{\ensuremath{\mathrm{\,Ge\kern -0.1em V}}\xspace}
\newcommand{\mev}{\ensuremath{\mathrm{\,Me\kern -0.1em V}}\xspace}
\newcommand{\mevcc}{\ensuremath{{\mathrm{\,Me\kern -0.1em V\!/}c^2}}\xspace}

\newcommand{\eff}{\varepsilon}

\def\pip        {\ensuremath{\pi^+}\xspace}
\def\pim        {\ensuremath{\pi^-}\xspace}

\def\fz#1       {\ensuremath{f_0({#1})}\xspace}

\def\BR{\mathcal{B}}

\usepackage{float}
\usepackage{lineno}
\usepackage{hyperref}

\hypersetup{colorlinks = true,
            linkcolor = blue,
            urlcolor = blue,
            citecolor = blue,
            breaklinks=true,
            pdfstartview=Fit}
\begin{document}

\newcommand{\BESIIIorcid}[1]{\href{https://orcid.org/#1}{\hspace*{0.1em}\raisebox{-0.45ex}{\includegraphics[width=1em]{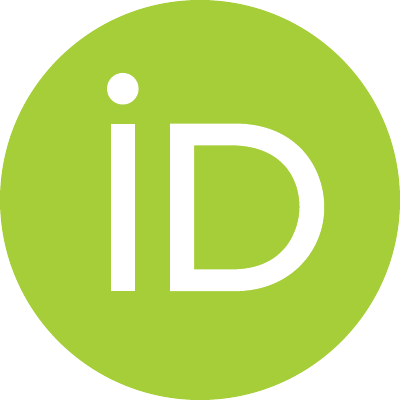}}}}

\title{\boldmath Search for the charged lepton flavor violating decay $\eta \to e^{\pm}\mu^{\mp}$}

\author{
\begin{center}
M.~Ablikim$^{1}$\BESIIIorcid{0000-0002-3935-619X},
M.~N.~Achasov$^{4,c}$\BESIIIorcid{0000-0002-9400-8622},
P.~Adlarson$^{83}$\BESIIIorcid{0000-0001-6280-3851},
X.~C.~Ai$^{89}$\BESIIIorcid{0000-0003-3856-2415},
C.~S.~Akondi$^{31A,31B}$\BESIIIorcid{0000-0001-6303-5217},
R.~Aliberti$^{39}$\BESIIIorcid{0000-0003-3500-4012},
A.~Amoroso$^{82A,82C}$\BESIIIorcid{0000-0002-3095-8610},
Q.~An$^{79,65,\dagger}$,
Y.~H.~An$^{89}$\BESIIIorcid{0009-0008-3419-0849},
Y.~Bai$^{63}$\BESIIIorcid{0000-0001-6593-5665}
% ...
\\
\vspace{0.2cm}
(BESIII Collaboration)\\
\vspace{0.2cm} {\it
$^{1}$ Institute of High Energy Physics, Beijing 100049, People's Republic of China\\
% ...
$^{4}$ Budker Institute of Nuclear Physics SB RAS (BINP), Novosibirsk 630090, Russia\\
% ...
$^{31}$ INFN Sezione di Ferrara, (A)INFN Sezione di Ferrara, I-44122, Ferrara, Italy; (B)University of Ferrara, I-44122, Ferrara, Italy\\
% ...
$^{39}$ Johannes Gutenberg University of Mainz, Johann-Joachim-Becher-Weg 45, D-55099 Mainz, Germany\\
% ...
$^{63}$ Southeast University, Nanjing 211100, People's Republic of China\\
% ...
$^{65}$ State Key Laboratory of Particle Detection and Electronics, Beijing 100049, Hefei 230026, People's Republic of China\\
% ...
$^{79}$ University of Science and Technology of China, Hefei 230026, People's Republic of China\\
% ...
$^{82}$ University of Turin and INFN, (A)University of Turin, I-10125, Turin, Italy; (B)University of Eastern Piedmont, I-15121, Alessandria, Italy; (C)INFN, I-10125, Turin, Italy\\
$^{83}$ Uppsala University, Box 516, SE-75120 Uppsala, Sweden\\
% ...
$^{89}$ Zhengzhou University, Zhengzhou 450001, People's Republic of China\\

\vspace{0.2cm}
$^{\dagger}$ Deceased\\
% ...
$^{c}$ Also at the Novosibirsk State University, Novosibirsk, 630090, Russia\\
% ...

}

\end{center}
}

\begin{abstract}

Based on $(10.087 \pm 0.044) \times 10^{9}~J/\psi$ events collected at the center-of-mass energy $\sqrt{s} = 3.097$~GeV with the BESIII detector, we search for the charged lepton flavor violating decay $\eta \to e^{\pm}\mu^{\mp}$ through the process $J/\psi \to \gamma \eta'$ with $\eta' \to \pi^{+} \pi^{-} \eta$. No signal is observed, and an upper limit on the branching fraction is determined to be $\mathcal{B}(\eta \to e^{\pm}\mu^{\mp}) < 6.8 \times 10^{-7}$ at the 90\% confidence level. This result improves the previous best limit by one order of magnitude.

\end{abstract}

\maketitle

\section{Introduction}

In the Standard Model (SM), charged lepton flavor violating (CLFV) processes are forbidden. Unlike other conservation laws in the SM, lepton flavor conservation is not associated with any conserved current derived from a fundamental symmetry. Therefore, numerous models beyond the SM naturally allow for CLFV processes, including the Minimal Supersymmetric extension of the SM~\cite{Brignole:2004ah}, Grand Unified Theories~\cite{Calibbi:2006nq}, and Seesaw Mechanisms~\cite{Ellis:2002fe}. Certain CLFV processes are predicted to occur at rates close to current experimental limits in specific parameter-dependent models~\cite{Hisano:1998fj,King:1998nv}.

Experimentally, the search for CLFV processes has a long history. Although no evidence has been observed so far, numerous upper limits have been established across a variety of channels, providing increasingly stringent constraints on theoretical models. To integrate different models, the effective field theory can be employed to describe CLFV processes. The CLFV decays of quarkonia, which possess different quantum numbers, show sensitivity to different Wilson coefficients. Therefore, by carefully selecting quarkonium states, we can access specific Wilson coefficients of interest~\cite{Hazard:2016fnc}. Extensive searches have been conducted on CLFV decays of quarkonia with various quantum numbers, including vector states such as \( \phi \)~\cite{Achasov:2009en}, 
\( J/\psi \)~\cite{BESIII:2013jau,BES:2004jiw}, and \( \Upsilon \)~\cite{CLEO:2008lxu,Belle:2022cce}, as well as pseudoscalar states like \( \pi^0 \)~\cite{Appel:2000wg,KTeV:2007cvy}, \( \eta \)~\cite{White:1995jc}, and \( \eta'(958) \)~\cite{CLEO:1999nsy}. However, no results are available for \( \omega \), \( \rho \), \( \eta_{c/b} \), or \( \chi_{c/bJ} \). Pseudoscalar states such as \( \eta \) and \( \eta' \) are exclusively sensitive to the axial coefficients, $C_{AL/AR}^{q\ell_{1}\ell_{2}}$, and the current upper limits on the branching fractions of the decays $\eta \to e^{\pm} \mu^{\mp}$ and $\eta' \to e^{\pm} \mu^{\mp}$ are $6.0 \times 10^{-6}$~\cite{White:1995jc} and $4.7 \times 10^{-4}$~\cite{CLEO:1999nsy}, respectively. The  $\BR(\eta \to e^\pm \mu^\mp)$ yields the most stringent bound $|C^{qe\mu}_{AL/AR}/\Lambda^2| < 3.0 \times 10^{-3}~(q=u,d)$ and $|C^{qe\mu}_{AL/AR}/\Lambda^2| < 2.1 \times 10^{-3}~(q=s)$, where $\Lambda$ denotes the energy scale of new physics and $AL/AR$ indicates the chirality of the operator.

The BESIII collaboration has collected the world's largest data samples of $(10.087 \pm 0.044) \times 10^{9}~J/\psi$ events produced in $e^{+} e^{-}$ annihilation. Given the large branching fractions $J/\psi \to \gamma \eta'$ and $\eta' \to \pip \pim \eta$, this dataset contains approximately $22.6 \times 10^{6}$ $\eta$ mesons, allowing to search for rare $\eta$ decays.

\section{BESIII DETECTOR AND MONTE CARLO SIMULATION}

The BESIII detector~\cite{BESIII:2009fln} records symmetric $e^+e^-$ collisions provided by the BEPCII storage ring~\cite{Yu:2016cof} in the center-of-mass energy range from 1.84 to 4.95~GeV, with a peak luminosity of $1.1 \times 10^{33}~\text{cm}^{-2}\text{s}^{-1}$ achieved at $\sqrt{s} = 3.773~\text{GeV}$. 
BESIII has collected large data samples in this energy region~\cite{BESIII:2020nme}. The cylindrical core of the BESIII detector covers 93\% of the full solid angle and consists of a helium-based
 multilayer drift chamber~(MDC), a time-of-flight
system~(TOF), and a CsI(Tl) electromagnetic calorimeter~(EMC),
which are all enclosed in a superconducting solenoidal magnet
providing a 1.0~T magnetic field.
The solenoid is supported by an octagonal flux-return yoke with resistive plate counter muon identification modules interleaved with steel. 
The charged-particle momentum resolution at $1~{\rm GeV}/c$ is $0.5\%$, and the ${\rm d}E/{\rm d}x$
resolution is $6\%$ for electrons from Bhabha scattering. The EMC measures photon energies with a resolution of $2.5\%$ ($5\%$) at $1$~GeV in the barrel (end cap) region. The time resolution in the plastic scintillator TOF barrel region is 68~ps, while that in the end cap region was 110~ps. The end cap TOF system was upgraded in 2015 using multigap resistive plate chamber technology, providing a time resolution of
60~ps, which benefits about 83\% of the data used in this analysis~\cite{Li:2017jpg}.

Simulated MC samples generated using a \textsc{geant4}-based~\cite{GEANT4:2002zbu} Monte Carlo (MC) package, which encompasses the geometric description of the BESIII detector and its response, are employed to determine detection efficiencies and estimate backgrounds. Approximately 10 billion $J/\psi$ inclusive MC events, corresponding to real data, are utilized to investigate backgrounds from $J/\psi$ decays. 

The process of $e^+e^-$ annihilation into a $J/\psi$ is modeled using \textsc{kkmc}~\cite{Jadach:2000ir}, which accounts for the beam energy spread and simulates initial state radiation. For the $J/\psi$ inclusive decays, \textsc{evtgen}~\cite{Lange:2001uf} is employed to simulate known decay modes, with branching fraction sourced from the Particle Data Group (PDG)~\cite{ParticleDataGroup:2024cfk}. For the remaining unknown decays, \textsc{lundcharm}~\cite{Chen:2000tv} is used. Additionally, final state radiation from charged final state particles is incorporated using the \textsc{photos} package~\cite{Barberio:1990ms}. To estimate the signal detection efficiency, 0.5 million signal MC events are generated. The process $J/\psi \to \gamma \eta'$ is simulated by \textsc{helamp} model~\cite{Lange:2001uf}, which allows simulation of any two-body decay by specifying the helicity amplitudes for the final-state particles. The subsequent process $\eta' \to \pi^{+} \pi^{-} \eta$ is generated by using a DIY generator based on the amplitude for $\eta' \to \pi^{+} \pi^{-} \eta$~\cite{BESIII:2017djm}, and $\eta \to e^{\pm} \mu^{\mp}$ is simulated by PHSP model~\cite{Lange:2001uf}, which assumes a uniform phase space distribution for the decay products.

To avoid involuntary bias during the analysis procedure, a semi-blind analysis method is used in this work. About 1.06 billion $J/\psi$ events (approximately 10\% of the full $J/\psi$ data) are used to validate the analysis strategy before unblinding. The full $J/\psi$ data sample is analyzed only after the analysis procedure has been fully fixed. In this paper, only results based on 
the full data sample are presented.

\section{Event selection and data analysis}
\label{sec:selection}

In this analysis, we investigate the decay $\eta \to e^{\pm} \mu^{\mp}$ through the decay chain $J/\psi \to \gamma \eta'$ with $\eta' \to \pi^{+} \pi^{-} \eta$. We require four charged tracks whose total charge is zero and at least one photon.
Charged tracks detected in the MDC are required to be within a polar angle ($\theta$) range of $|\rm{cos\theta}|<0.93$, where $\theta$ is defined with respect to the $z$-axis,
which is the symmetry axis of the MDC. For charged tracks, the distance of closest approach to the interaction point (IP) 
must be less than 10\,cm
along the $z$-axis,  
and less than 1\,cm
in the transverse plane.

Photon candidates are identified using isolated showers in the EMC.  The deposited energy of each shower must be more than 25~MeV in the barrel region ($|\cos \theta|< 0.80$) and more than 50~MeV in the end cap region ($0.86 <|\cos \theta|< 0.92$).  To exclude showers that originate from charged tracks,
the angle subtended by the EMC shower and the position of the closest charged track at the EMC must be greater than 10 degrees as measured from the IP. 
To suppress electronic noise and showers unrelated to the event, the difference between the EMC time and the event start time is required to be within 
[0, 700]\,ns.

The final states are composed of $\pi^{+}$, $\pi^{-}$, $e^{\pm}$, $\mu^{\mp}$, and $\gamma$. In this decay, the momenta of $\mu^{\pm}$ and $\pi^{\pm}$ are low and largely overlapping, making it difficult to distinguish muons from pions. Therefore, we apply a combination of particle identification (PID) and the four-constraints (4C) kinematic fit to separate the charged particles $\pi^{+}$, $\pi^{-}$, $e^{\pm}$, and $\mu^{\mp}$. The PID for each charged track hypothesis is based on the $dE/dx$, EMC and TOF information and the $\chi^2(i)=\chi^2_{dE/dx}+\chi^2_{\rm TOF}+\chi^2_{\rm EMC}$ is calculated. The combined $\chi^{2}_{\rm PID}(\pi^{+} \pi^{-} e^{\pm} \mu^{\mp}) =  \chi^{2}(\pi^{+}) + \chi^{2}(\pi^{-}) + \chi^{2}(e^{\pm})+ \chi^{2}(\mu^{\mp})$ value is determined for all the track hypotheses. Then, 4C kinematic fits are performed under all different photon combinations and charge track hypotheses, which enforce energy-momentum conservation. The combination with the minimum $\chi^{2}_{\rm 4C+PID}(\pi^{+} \pi^{-} e^{\pm} \mu^{\mp}) = \chi^{2}_{\rm 4C}(\gamma \pi^{+} \pi^{-} e^{\pm} \mu^{\mp}) + \chi^{2}_{\rm PID}(\pi^{+} \pi^{-} e^{\pm} \mu^{\mp})$ is retained and required to be less than 36, which is optimized by the Punzi Figure of Merit (FOM)~\cite{Punzi:2003bu}  
\begin{equation}
\rm FOM = \frac{\eff_{\rm signal}}{a/2 + \sqrt{B}},
\end{equation}
where $\eff_{\rm signal}$ represents the detection efficiency from MC simulation, a is the significance value set to 3, and $B$ is the expected number of background events, which is estimated using the $J/\psi$ inclusive MC samples.

Additionally, the $\chi^{2}_{\rm 4C+PID}$ values under the $\pip \pim e^{+}e^{-}$ and $\pip \pim \pip \pim$ hypotheses are also calculated to further reduce background. To suppress these background\hyp{}s, $\chi^{2}_{\rm 4C+PID}(\pi^{+} \pi^{-} e^{\pm} \mu^{\mp})<\chi^{2}_{\rm 4C+PID}(\pi^{+} \pi^{-} e^{+} e^{-})$ and $\chi^{2}_{\rm 4C+PID}(\pi^{+} \pi^{-} e^{\pm} \mu^{\mp})<\chi^{2}_{\rm 4C+PID}(\pi^{+} \pi^{-} \pi^{+} \pi^{-})$ are also required for candidate events. 
Further requirements on the ratio between energy deposit in the EMC and momentum is required to be larger than 0.8 for electron. 
And the deposited energy for muon is required to be in (0.1, 0.3) GeV.

\begin{figure}[H]
    \centering
    \includegraphics[width=0.48\textwidth]{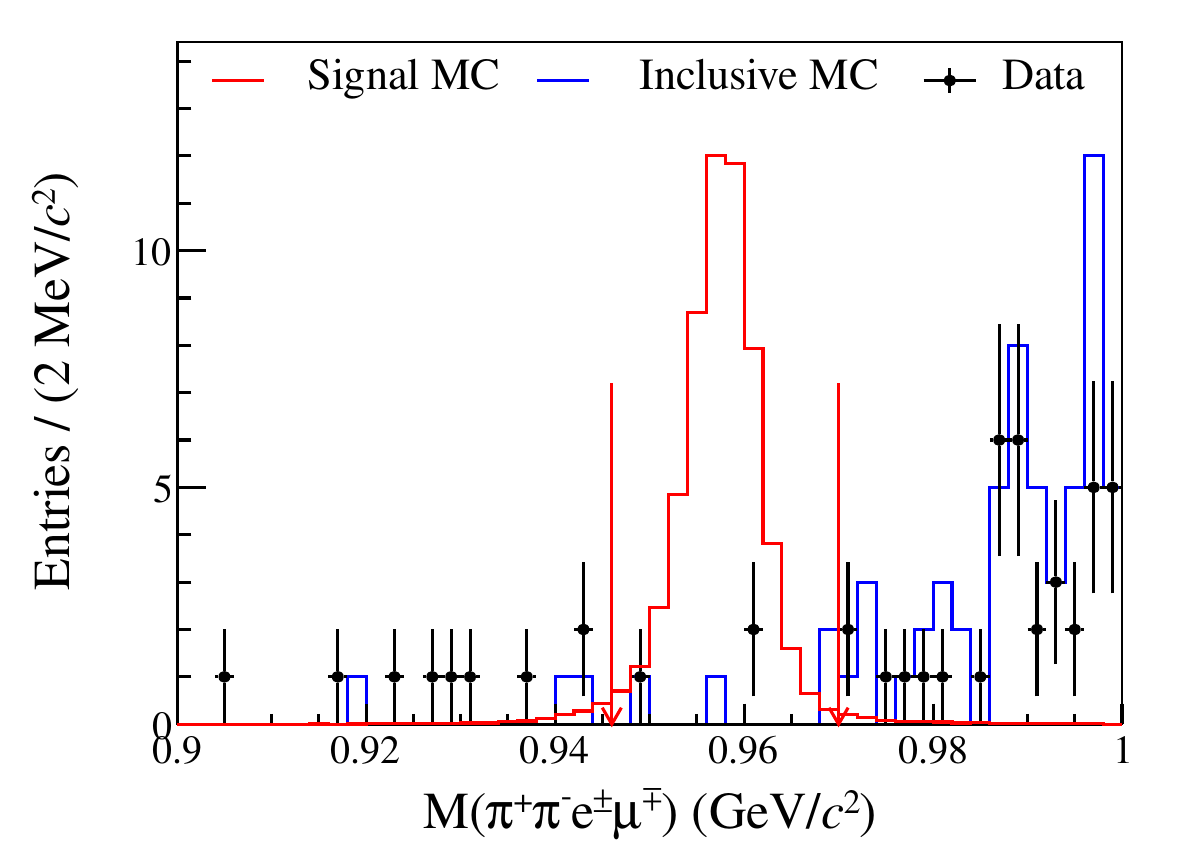}
    \caption{The distribution of $M(\pi^{+} \pi^{-} e^{\pm} \mu^{\mp})$. The red histogram illustrates the signal MC sample, the blue histogram represents the inclusive MC sample, and the black dots with error bars denote the data. The red arrows indicate the signal range, $M(\pi^{+} \pi^{-} e^{\pm} \mu^{\mp}) \in (0.946, 0.97)~\text{GeV}/c^2$.}
       \label{fig:Mpipiemu}
\end{figure}

After applying the above selection criteria, the invariant mass of $\pi^{+} \pi^{-} e^{\pm} \mu^{\mp}$, $M(\pi^{+} \pi^{-} e^{\pm} \mu^{\mp})$, is shown in Fig.~\ref{fig:Mpipiemu}. To further reduce backgrounds that do not come from the $\eta'$, $M(\pi^{+} \pi^{-} e^{\pm} \mu^{\mp})$ is required to be in the $\eta'$ signal region $(0.946, 0.97)~\text{GeV}/c^2$, which is around three times the $\eta'$ mass resolution.

\begin{figure}
\centering
        \centering
        \includegraphics[width=0.48\textwidth]{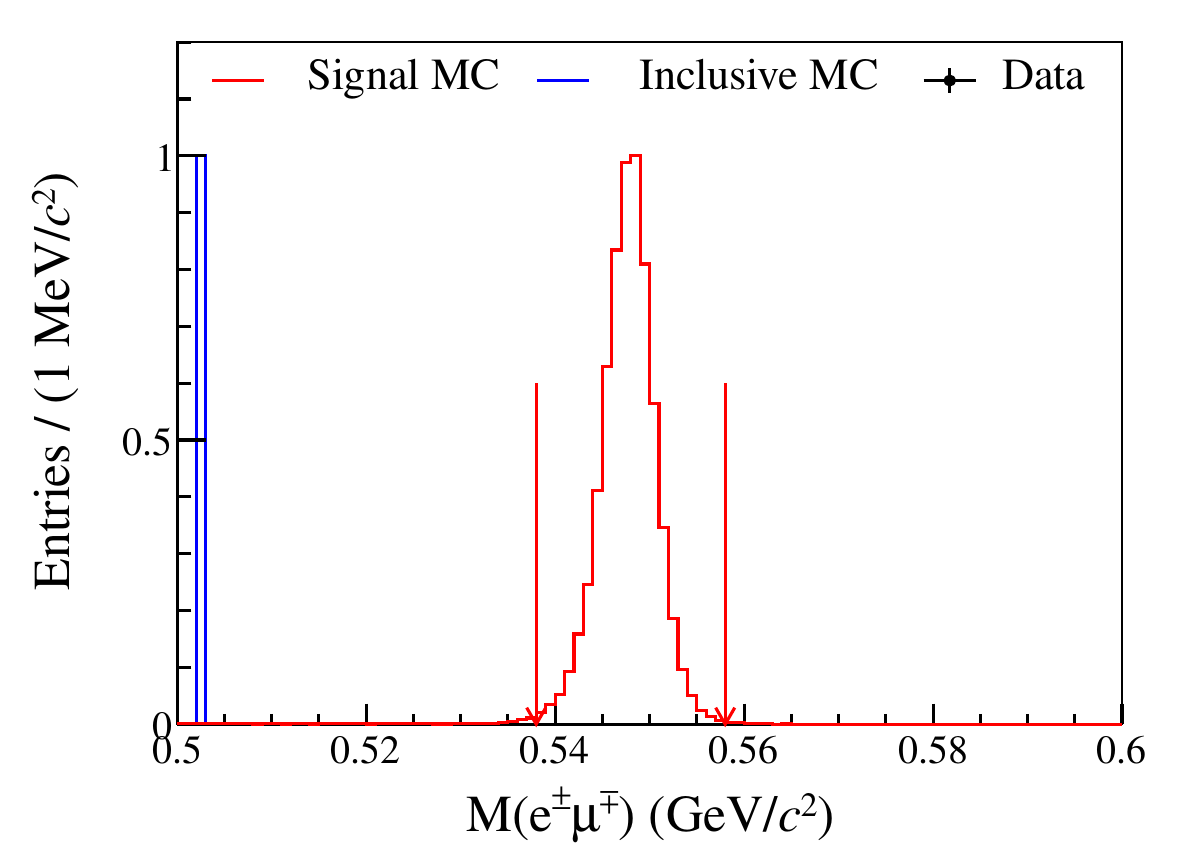}
\caption{The distribution of $M(e^{\pm}\mu^{\mp})$. The red histo\hyp{}gram illustrates the signal MC sample, the blue histogram represents the inclusive MC sample, and no data survived. The red arrows indicate the signal range, $M(e^{\pm} \mu^{\mp}) \in (0.538, 0.558)~\text{GeV}/c^2$.}
\label{fig:memu_2}
\end{figure}

With the selection criteria applied, the signal region of the invariant mass of $e^{\pm} \mu^{\mp}$, $M(e^{\pm} \mu^{\mp})$, is defined as $(0.538,0.558)~\text{GeV}/c^2$, corresponding to three times the $\eta$ mass resolution. No events from either the data or the inclusive MC sample are observed in the $\eta$ signal region, as shown in Fig.~\ref{fig:memu_2}.

To further check the continuum background contribution, data samples collected at center-of-mass energies ($\sqrt{s_i}$) of 3.08~GeV, 3.65~GeV, 3.682~GeV, and 3.773~GeV, corresponding to integrated luminosities ($\mathcal{L}_i$) of $166.3~\rm pb^{-1}$, $455~\rm pb^{-1}$, $404~\rm pb^{-1}$, and $20274.8~\rm pb^{-1}$, were used and no events survived the above selection criteria. With a dedicated signal MC sample, the selection efficiency is estimated to be $(14.96 \pm 0.05) \%$ with statistical uncertainty only.

\section{SYSTEMATIC UNCERTAINTY}
\label{sec:systematics}
The systematic uncertainties in the measurement of the branching fraction for $\eta \rightarrow e^{\pm}\mu^{\mp}$ primarily arise from several sources: the total number of $J/\psi$ events, the branching fractions of decays of intermediate states, tracking, PID, photon selection, and 4C kinematic fit. The systematic uncertainties from these sources are investigated and discussed below.

The number of $J/\psi$ events in data is determined to be $(10.087 \pm 0.044) \times 10^{9}$, and its uncertainty is estimated to be 0.5\%~\cite{BESIII:2021cxx}. The world-average~\cite{ParticleDataGroup:2024cfk} branching fractions of the $J/\psi \to \gamma \eta'$ and $\eta' \to \pi \pi \eta$ decays are used, which both have 1.2\% relative uncertainty.

The difference between data and MC samples for the tracking efficiency of charged pions is investigated using a control sample of $J/\psi \rightarrow \pi^{+}\pi^{-}\pi^{0}$ events~\cite{Liu:2024uno}. For the tracking of electrons and positrons, the radiative Bhabha process $e^{+}e^{-} \to e^{+}e^{-} \gamma$ from the data samples collected at the center-of-mass energies of 3.08 GeV and 3.097 GeV is used~\cite{Chai:2025xni}. The systematic uncertainty due to the tracking of muons has been assigned as 1.0\% per track by using the control sample of $e^{+} e^{-} \to \gamma \mu^{+} \mu^{-} $ in Ref.~\cite{BESIII:2022exh}. A transverse momentum and $\cos \theta$ dependent correction to the detection efficiency is obtained by comparing the efficiency between the data and MC simulation, where $\theta$ is the polar angle of the track. The differences in the selection efficiencies are taken as the systematic uncertainties, which are 0.4\% for $\pi^{+}$, 0.2\% for $\pi^{-}$, 1.2\% for $e^{\pm}$, and 1.0\% for $\mu^{\mp}$. The total uncertainty is calculated as $(1-(1-0.4\%) (1-0.2\%) (1-1.2\%) (1-1.0\%) = 2.8\%)$, which is approximately the linear sum. The  uncertainty associated with tracking of charged particles amounts to 2.8\% in total.

The systematic uncertainty associated with PID mainly originates from two sources: one comes from the $\chi^2_{\rm PID}$ information used in the combined PID and 4C kinematic fit, and the other part arises from the further requirements for electron and muon.
The uncertainties associated with the first part are evaluated using the same method as that employed for tracking efficiency studies, which are estimated to be 0.3\% for $\pi^{+}$, 0.3\% for $\pi^{-}$, 0.5\% for $e^{\pm}$, and 1.0\% for $\mu^{\mp}$. 
A mixed sample of $e^{+} e^{-} \to \gamma e^{+} e^{-} $ at the $J/\psi$ meson mass and $J/\psi \rightarrow \gamma e^{+} e^{-} $~\cite{Chai:2025xni} is used to estimate the uncertainty associated with the additional requirement for electron, which is estimated to be 5.8\%. A control sample for $J/\psi \to \gamma \mu^{+} \mu^{-}$ is used to estimate the extra uncertainty for muon selection, which is estimated to be , 3.1\%. The overall uncertainty associated with PID amounts to 6.9\%.

The systematic uncertainty of photon detection is investigated using the control sample of $e^{+}e^{-} \to \gamma \mu^{+} \mu^{-}$~\cite{BESIII:2015rug}. The results indicate that the discrepancy between data and MC simulation is approximately 0.5\%, which is taken as the systematic uncertainty per photon.

The uncertainty in the 4C kinematic fit arises from the difference between data and MC simulation. This difference is reduced by correcting the track helix parameters in the MC simulation~\cite{BESIII:2012mpj}, which is found to be  2.2\%.

All the systematic uncertainties discussed above are summarized in Tab.~\ref{tab:systematics}, where the total systematic uncertainty, 8.0\%, is given by the quadratic sum of the individual uncertainties, assuming all the sources are independent.
\begin{table}[ht]
\centering
\caption{Relative systematic uncertainties for the $\eta \to e^{\pm} \mu^{\mp}$ ($J/\psi\to\gamma\eta^{\prime}$, $\eta^{\prime}\to\pi^{+}\pi^{-}\eta$) process.}
\label{tab:systematics}
\begin{tabular}{ccc}
\hline \hline
Source & Uncertainty (\%) \\
\hline
Number of $J/\psi$ events & 0.5 \\
$\BR(J/\psi\to\gamma\eta^{\prime})$ & 1.2 \\
$\BR(\eta^{\prime}\to\pi^{+}\pi^{-}\eta)$ & 1.2 \\
Tracking & 2.8 \\
PID & 6.9 \\
Photon detection & 0.5 \\
4C kinematic fit & 2.2 \\
\hline
Total & 8.0 \\
\hline \hline
\end{tabular}
\end{table}

\section{Result}
No signal events are observed in the signal region, while the number of background events is expected to be 0. Hence, no excess is observed and an upper limit on the branching fraction $\BR(\eta \to e^{\pm}\mu^{\mp})$ is estimated with the profile likelihood method~\cite{Rolke:2004mj}. The likelihood function is defined as
\begin{equation}
\begin{split}
\mathcal{L}(\BR, \eff_{\rm sig}, N_{\rm bkg}) =& \mathcal{P}\left(N_{\rm obs} \mid N_{\eta} \cdot \BR \cdot \eff_{\rm sig} + N_{\rm bkg} \right) \\
&\cdot \mathcal{P}\left(N_{\rm bkg}^{\rm inclmc} \mid N_{\rm bkg}\right) \\
&\cdot \mathcal{G}\left(\eff_{\rm sig} \mid \eff_{\rm sig}^{\rm MC}, \eff_{\rm sig}^{\rm MC} \cdot \sigma_{\rm sys}\right),
\end{split}
\label{eq:likelihood function}
\end{equation}
where the number of observed events, $N_{\rm obs}$, is assumed to follow a Poisson distribution $(\mathcal{P})$ with mean equal to the sum of the expected signal yield from the $\eta$ sample $(N_{\eta})$ and background events $(N_{\rm bkg})$; $N_{\eta}$ is equal to $N_{J/\psi} \cdot \BR(J/\psi \to \gamma \eta') \cdot \BR(\eta' \to \pip \pim \eta)$, where $N_{J/\psi}$ is the number of $J/\psi$ events in the data. Given the limited number of events that survive in the data samples for the background study, the background events are assumed to follow a Poisson distribution $\mathcal{P}$. For the efficiency of signal detection, we assume it follows a Gaussian distribution $\mathcal{G}$ with the mean and width corresponding to the MC-determined nominal efficiency $\eff_{\rm sig}^{\rm MC}$ and its absolute uncertainty, $\eff_{\rm sig}^{\rm MC} \cdot \sigma_{\rm sys}$, with the systematic uncertainty ($\sigma_{\rm sys}$) considered. The likelihood distribution serves as a probability function for the signal branching fraction $\BR$. The fixed parameters in the likelihood function are $N_{\eta}$, $N^{\rm inclmc}_{\rm bkg}$, $\eff^{\rm MC}_{\rm sig}$, $\sigma_{\rm sys}$, with corresponding values of $2.26 \times 10^{7}$, 0, 14.96\% and 8.0\%, respectively.

\begin{figure}[H]
    \centering
    \includegraphics[width=0.99\linewidth]{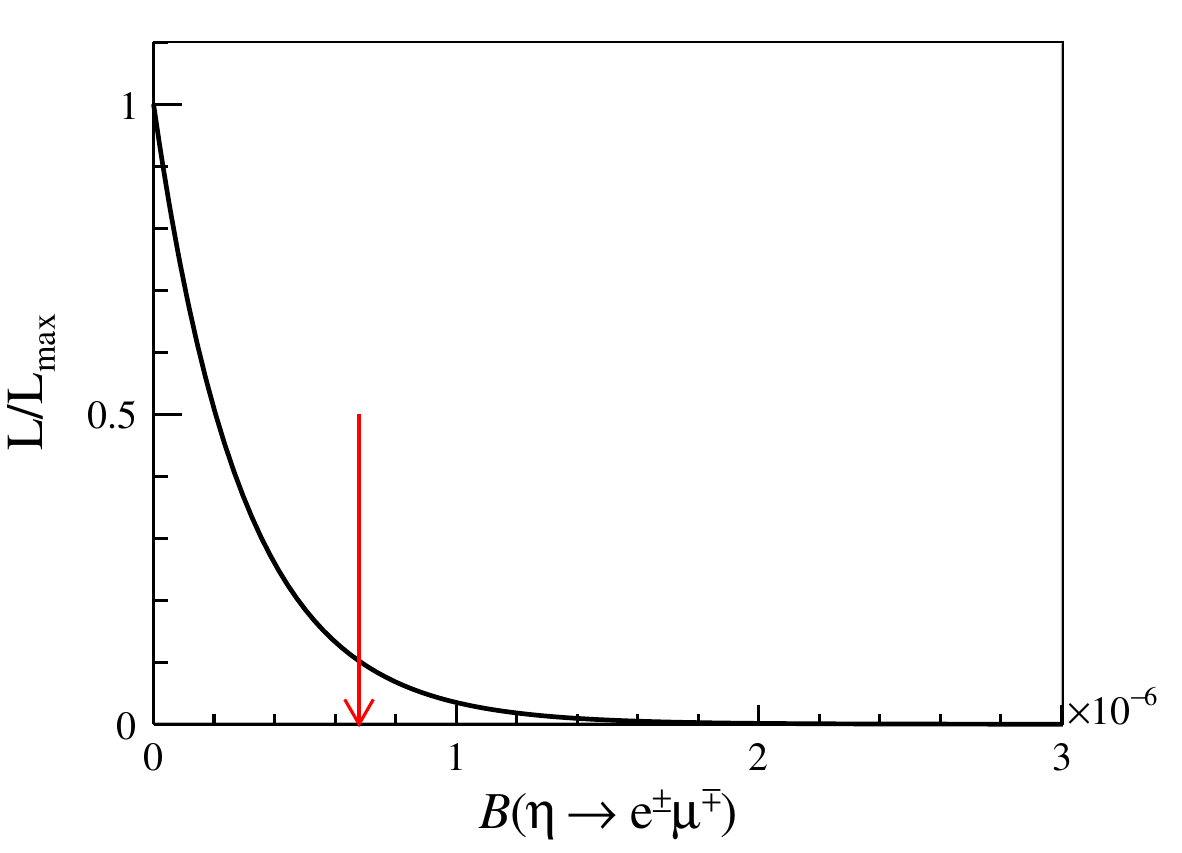}
    \caption{The profile likelihood distribution as a function of $\BR(\eta \rightarrow e^{\pm}\mu^{\mp})$. The red arrow denotes the upper limit of branching fraction at the 90\% C.L..}
    \label{fig:BF}
\end{figure}

By scanning the number of signal events, a likelihood distribution is obtained, resulting in a set of likelihood values
$\mathcal{L}_{i}$, with a maximum value $\mathcal{L}_{\rm max}$. The normalized likelihood, defined as $\mathcal{L}_{i}/\mathcal{L}_{\rm max}$, versus the branching fraction is shown in Fig.~\ref{fig:BF}. By integrating the likelihood distribution to 90\% of the total probability, the upper limit on the $\eta \rightarrow e^{\pm}\mu^{\mp}$ branching fraction is determined to be $6.8 \times 10^{-7}$ at the 90\% confidence level (C.L.).

According to Ref.~\cite{Hazard:2016fnc}, for pseudoscalar mesons, the constraints on Wilson coefficients are in direct proportion to the square root of the upper limit on the branching fraction
\begin{equation}
\label{eq11}
\left|C/\Lambda^{2}\right| \propto \sqrt{\BR(P \rightarrow \ell_{1}^{\pm} \ell_{2}^{\mp})},
\end{equation}
where $C$ denotes $C_{AL/AR}^{q\ell_{1}\ell_{2}}$, $C_{PL/PR}^{q\ell_{1}\ell_{2}}$, or $C_{\tilde{G}L/\tilde{G}R}^{\ell_{1}\ell_{2}}$, which are the left- or right-handed Wilson coefficients of the axial-vector, pseudoscalar, and gluonic operators, respectively. It is important to note that the contributions from other Wilson coefficients are neglected when imposing a constraint on one Wilson coefficient. The obtained constraints are summarized in Tab.~\ref{tab:new_wilson_coefficients}. 
Compared with the upper limit $\mathcal{B}(\eta \to e^{\pm} \mu^{\mp}) < 6.0 \times 10^{-6}$ from the PDG~\cite{ParticleDataGroup:2024cfk}, this result leads to a strengthened constraints on the Wilson coefficients.

\begin{table}[ht]
\centering
\renewcommand{\arraystretch}{2}
\caption{Updated constraints on the Wilson coefficients from $\eta$ meson decays.}
\label{tab:new_wilson_coefficients}
\begin{tabular}{ccccc}
\hline \hline
\multirow{2}{*}{\makecell{Wilson\\coefficients}}  & \multicolumn{2}{c}{PDG} & \multicolumn{2}{c}{This work} \\
\cmidrule(lr){2-3} \cmidrule(lr){4-5}
 (GeV$^{-2}$)  & $\eta(u/d)$  & $\eta(s)$ & $\eta(u/d)$ & $\eta(s)$ \\
\midrule
$\left|C_{AL/AR}^{qe\mu}/\Lambda^{2}\right|$  & $3.0 \times 10^{-3}$ & $2.1 \times 10^{-3}$ & $1.0 \times 10^{-3}$ & $7.0 \times 10^{-4}$ \\
$\left|C_{PL/PR}^{qe\mu}/\Lambda^{2}\right|$  & $1.7 \times 10^{3}$ & $1.2 \times 10^{3}$ & $5.8 \times 10^{2}$ & $4.0 \times 10^{2}$ \\
$\left|C_{\tilde{G}L/\tilde{G}R}^{e\mu}/\Lambda^{2}\right|$  & \multicolumn{2}{c}{$1.6 \times 10^{2}$} & \multicolumn{2}{c}{$5.4 \times 10^{1}$}  \\
\hline \hline
\end{tabular}
\end{table}

\section{Summary}
The large statistics of $J/\psi$ data provides an opportunity to investigate the CLFV decay $\eta \to e^{\pm} \mu^{\mp}$, which would serve as a clear signature of new physics. Using $(10.087 \pm 0.044) \times 10^{9} J/\psi$ events collected with the BESIII detector at the BEPCII, no events are observed within the signal window of the data sample. The upper limit on $\BR(\eta \rightarrow e^{\pm}\mu^{\mp})$ via $J/\psi \to \gamma \eta'$ and $\eta' \to \pi^{+} \pi^{-} \eta$ is determined to be $6.8 \times 10^{-7}$ at the 90\% C.L., enhancing the precision by one order of magnitude compared to the PDG value of $6.0 \times 10^{-6}$. Consequently, the upper limits on the Wilson coefficients, $C_{AL/AR}^{qe\mu}$, $C_{PL/PR}^{qe\mu}$, and $C_{\tilde{G}L/\tilde{G}R}^{e\mu}$, are further constrained.

\section{ACKNOWLEDGMENTS}

The BESIII Collaboration thanks the staff of BEPCII (https://cstr.cn/31109.02.BEPC) and the IHEP computing center for their strong support. This work is supported in part by National Key R\&D Program of China under Contracts Nos. 2023YFA1606000, 2023YFA1606704; National Natural Science Foundation of China (NSFC) under Contracts Nos. 11635010, 11935015, 11935016, 11935018, 12025502, 12035009, 12035013, 12061131003, 12192260, 12192261, 12192262, 12192263, 12192264, 12192265, 12221005, 12225509, 12235017, 12342502, 12361141819; the Chinese Academy of Sciences (CAS) Large-Scale Scientific Facility Program; the Strategic Priority Research Program of Chinese Academy of Sciences under Contract No. XDA0480600; CAS under Contract No. YSBR-101; 100 Talents Program of CAS; The Institute of Nuclear and Particle Physics (INPAC) and Shanghai Key Laboratory for Particle Physics and Cosmology; ERC under Contract No. 758462; German Research Foundation DFG under Contract No. FOR5327; Istituto Nazionale di Fisica Nucleare, Italy; Knut and Alice Wallenberg Foundation under Contracts Nos. 2021.0174, 2021.0299, 2023.0315; Ministry of Development of Turkey under Contract No. DPT2006K-120470; National Research Foundation of Korea under Contract No. NRF-2022R1A2C1092335; National Science and Technology fund of Mongolia; Polish National Science Centre under Contract No. 2024/53/B/ST2/00975; STFC (United Kingdom); Swedish Research Council under Contract No. 2019.04595; U. S. Department of Energy under Contract No. DE-FG02-05ER41374; Guangdong Basic and Applied Basic Research Foundation 2024A1515012416



\bibliographystyle{apsrev4-2}
\bibliography{pipiemu-draft}

\end{document}